\documentclass[letterpaper,conference]{IEEEtran}

\usepackage[compress]{cite}
\usepackage{subfigure} 
\usepackage{amsmath,amssymb,amsfonts,amsthm}
\usepackage{textcomp}
\usepackage{xcolor}
\usepackage{soul} 
\usepackage{booktabs} 
\usepackage{makecell} 
\usepackage{pgf,pgfkeys,tikz,circuitikz,float,bm,caption}
\usepackage[ruled,vlined]{algorithm2e}
\usepackage{graphicx,multirow}%
\usepackage[super]{nth}
\usepackage{steinmetz}
\usepackage{siunitx,upgreek}
\usepackage{orcidlink}
\def\BibTeX{{\rm B\kern-.05em{\sc i\kern-.025em b}\kern-.08em
		T\kern-.1667em\lower.7ex\hbox{E}\kern-.125emX}}

\DeclareSIUnit{\rad}{rad}

\graphicspath{{figures/}}

\graphicspath{{figures/}}

\begin{document}
	
	\title{The micro-Doppler Attack Against AI-based Human Activity Classification from Wireless Signals}

		%
	
	\author{ 
		\IEEEauthorblockN{Margarita Loupa\IEEEauthorrefmark{1}, Antonios Argyriou\IEEEauthorrefmark{1}, Yanwei Liu\IEEEauthorrefmark{2}}
		\IEEEauthorblockA{\IEEEauthorrefmark{1}Department of Electrical and Computer Engineering, University of Thessaly, 38334 Volos, Greece.\\\IEEEauthorrefmark{2}Institute of Information Engineering, Chinese Academy of Sciences, Beijing 100093, China.
		}
	}



\maketitle

\IEEEpubidadjcol

\begin{abstract}
	A subset of Human Activity Classification (HAC) systems are based on AI algorithms that use passively collected wireless signals. This paper presents the \textit{micro-Doppler attack} targeting HAC from wireless orthogonal frequency division multiplexing (OFDM) signals. The attack is executed by inserting artificial variations in a transmitted OFDM waveform to alter its micro-Doppler signature when it reflects off a human target. We investigate two variants of our scheme that manipulate the waveform at different time scales resulting in altered receiver spectrograms.  HAC accuracy with a deep convolutional neural network (CNN) can be reduced to less than 10\%. 
	
\end{abstract}


\section{Introduction} 
WiFi signals have been used successfully for passive tracking, localization, and human activity classification (HAC) ~\cite{Berger10,wisee13,WiVi13,WiHear14,Wang14,Chen15}. These systems leverage the micro-Doppler effects that alter the signals that impinge on humans. Unique features in the spectrogram due to the movement of various body parts allow the classification of activities. Most of these HAC systems strive to use opportunistically received signals from any neighboring wireless transmitter. What we argue in this paper is that these passive systems, although extremely useful, may be vulnerable to attacks from the emitter of the RF signal. 
%
%
To demonstrate that, we propose a novel \textit{micro-Doppler attack} that employs \textit{waveform modification} at the transmitter to prevent correct micro-Doppler estimation and passive activity classification. We focus on orthogonal frequency division multiplexing (OFDM) waveforms since they are widely used in wireless local area networks (WLANs) like WiFi, and also cellular communication standards. 

\begin{figure}[t]
	\centering
	\includegraphics[width=0.85\linewidth]{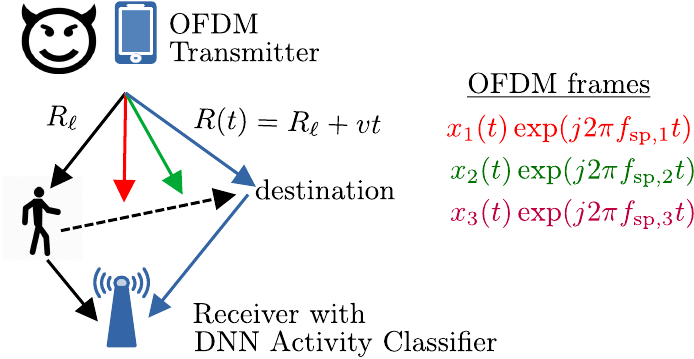}
	\caption{A human moves with a velocity $v$ that generates a change in the length of the wireless transmission path from the Tx to the Rx equivalent to $vt$. Different color indicates different frame, and each frame with different artificial Doppler.}
	\label{fig:topology-reflections}
\end{figure}

Fig.~\ref{fig:topology-reflections} illustrates a representative scenario where a WiFi transmitter (Tx) emits successively OFDM frames illustrated with different colors. The human is located at an initial bistatic range $R_l$ which is the distance from Tx to the human and from the human to the Rx. This bistatic range changes to $R(t)$ as the human moves with speed $v$ to the destination. In this paper, we alter the transmitted OFDM signal, denoted as $x(t)$, by inserting artificial frequency and phase shifts across frames to prevent a correct spectrogram estimation that reveals the unique features of human activities~\cite{wisee13,Wang14}.

There is little focus on attacking these passive HAC systems. Most recent works on signal modification focus on preventing signal classification as a whole, rather than specific features that enable HAC. The authors in~\cite{Rahbari14} obfuscate the modulation type of the wireless signal but this does not affect the micro-Doppler which is key in HAC. This method, however, was not intended by its design to prevent the estimation of micro-Doppler signatures of opportunistically illuminated targets. 
In~\cite{jnl_2023_access} it was shown that mixing a frequency-modulated (FM) signal with modulated information symbols at a transmitter will result in a smeared spectrogram at a receiver. This FM-based technique was used for attacking AI-enabled classifiers, namely a Deep Neural Network (DNN), but to hide the wireless modulation type~\cite{varkatzas2023}. A closely related work to these papers is~\cite{PhyCloak16} which distorts Doppler information through a third relay node that emits simultaneously with the transmitter. 
The previous methods deal with micro-Doppler signatures of opportunistically illuminated targets (like Fig.~\ref{fig:topology-reflections}). 
%
%
Unlike the literature, the proposed \textit{micro-Doppler attack} is carried out with a waveform that is designed to prevent the creation of meaningful spectrograms when a wireless signal is reflected from targets, thus leading to poor performance of an AI-enabled HAC system. 

\section{Signal Models and the micro-Doppler Attack}
\label{section:signal-model}

\subsection{Channel and Signal Model}
The OFDM transmitter emits frames (different colors in Fig.~\ref{fig:topology-reflections} indicate a different frame), where each one consists of several OFDM symbols. We assume that the channel formed between the Tx-target-Rx is constant for the duration of the frame due to the slow velocity of human movement. We index the OFDM frames with the letter $m$. 
%
%
%
%
%
%
%
The analog baseband transmitted signal is $x(t)$ and is received over multiple paths that are created from different body parts. When the human moves, the length of each one of these paths changes. If $R_l$ is the length of the initial path, with movement the transmission delay for the $l$-th path is $\tau_l(t)$=$\frac{R(t)}{c}$=$\frac{R_l}{c}-\frac{v_l}{c}t$. $f_c$ is the carrier frequency and so $f_{D_l}$=$f_c\frac{v_l}{c}$ is the Doppler frequency shift. Fig.~\ref{fig:topology-reflections} illustrates this change in the path length due to human movement. The received baseband signal is: 
\begin{align}
	y(t)=\sum_{l} a'_l x(t-\tau_l(t))e^{-j2\pi f_c\tau_l(t)}+w(t).
	\label{eqn:signal-model-analog}
\end{align}
In the above $a'_l$ is the complex amplitude of the $l$-th path that includes also distance-dependent path loss. $\sigma^2$ is the variance of the receiver additive white Gaussian noise (AWGN) $w(t)$. 

We expand the previous model when $x(t)$ is the result of OFDM. With $N$ sub-carriers that can contain data, pilot symbols, or a combination of both (depending on the standard), the desired OFDM symbol in continuous time is:
\begin{equation}
	x(t)=\frac{1}{\sqrt{N}}\sum_{k=0}^{N-1}\tilde{x}[k]e^{j2\pi f_k t},~~0\leq t \leq T_N.
	\label{eqn:ct-ofdm} 
\end{equation}
$\tilde{x}[k]$ is the complex symbol modulated onto subcarrier $k$, and if subcarrier spacing is $\Delta f$, then $T_N$=$N/\Delta f$ is the OFDM symbol duration. Substituting \eqref{eqn:ct-ofdm} into \eqref{eqn:signal-model-analog}:
\begin{align*}
	&y(t)= \frac{\sum_l a'_l}{\sqrt{N}}\sum_{k=0}^{N-1}\tilde{x}[k]e^{j2\pi f_k( t-\frac{R_l}{c})} e^{-j2\pi f_c(\frac{R_l}{c}-\frac{v_lt}{c})}+w[n].
\end{align*}
To get the sampled version of the last expression, we first set $a_l$=$a'_le^{-2\pi f_c\frac{R_l}{c}}$. Then we sample at times $t=n/f_s+mT_N$ to obtain the data cube that contains the fast (index $n$) and slow (index $m$) time samples (see~\cite{cnf_2023_radarconf1} for details): \normalsize
\begin{align}
	y[n,m]= \frac{\sum_l a_l}{\sqrt{N}}\sum_{k=0}^{N-1}\tilde{x}[k]e^{j2\pi \big ( f_k( \frac{n}{f_s}-\frac{R_l}{c})+f_{D_l}mT_N \big )} +w[n].	\label{eqn:reflected-signal-model-discrete-1}
\end{align}
\normalsize
This last expression came from noticing that $e^{j2\pi f_k mT_N}=1$, since $f_kT_N$= $k\Delta f T_N$=$kN$. It also includes the assumption that the Doppler within one OFDM symbol is negligible so $e^{-j2\pi f_{D_l}\frac{n}{f_s}} \approx 1$. This is valid due the small speed of a moving human and all its body parts~\cite{Boulic90,Vishwakarma21}.

We want to develop the last model into a more compact vector form so that additional effects like multipath channels and receiver processing can be understood more clearly.
To do that, first the Quadrature Amplitude Modulated (QAM) symbols are compactly presented as $\tilde{\mathbf{x}}=[\tilde{x}[1],...,\tilde{x}[k],..,\tilde{x}[N]]$. The Doppler impact due to the potential carrier frequency offset (CFO) is $\exp (j 2 \pi f_\text{cfo} t)$ and can be captured as the diagonal matrix $[\mathbf{C}_\text{cfo}]_{n,n}$=$\exp (j 2 \pi f_\text{cfo} n/f_s)$. However, the Doppler of the $l$-th path in ~\eqref{eqn:reflected-signal-model-discrete-1} is the single constant complex number $c_{\text{ch},l}$=$e^{j2\pi f_{D_l}mT_N}$ that changes only across OFDM frames indexed by $m$.
The impact of the path distance $R_l$ on each subcarrier $k$ is accounted for in the diagonal matrix $\mathbf{P}_\text{ch}$ where the $k$-th diagonal element is $
[\mathbf{P}_\text{ch}]_{k,k}$=$\exp (-j 2 \pi f_k \frac{R_l}{c})$, $k$=$1,...,N$. 
Finally, with $\mathbf{F^{H}}$ we indicate the $N\times N$ inverse discrete Fourier Transform (IDFT) matrix whose $n,k$ elements are (recall that in OFDM the subcarriers are placed at the orthogonal IDFT frequencies so $f_k=k\Delta f =kf_s/N$): 
\begin{align}
	[\mathbf{F^{H}}]_{n,k}=e^{j2\pi nk/N}=e^{j2\pi nk\Delta f/f_s},0\leq k,n\leq N.
\end{align}
We vectorize~\eqref{eqn:reflected-signal-model-discrete-1} and obtain the time domain signal over the $N$ samples of a single OFDM symbol in the $m$-th OFDM frame:
\begin{align}
	\mathbf{y}=\mathbf{C}_\text{cfo}\mathbf{F^{H}} \Big ( \sum_l a_l c_{\text{ch},l}\mathbf{P}_{\text{ch},l} \Big )\tilde{\mathbf{x}}+\tilde{\mathbf{w}}.
\end{align}
The receiver collects $\mathbf{y}$ for each one of the OFDM symbols in the preamble of a frame that contains known symbols in $\tilde{\mathbf{x}}$.

\subsection{The micro-Doppler Attack with OFDM Pre-Coding}
To alter the micro-Doppler signature we multiply each one of the OFDM symbols of the $m$-th frame with the diagonal pre-coding matrix $\mathbf{P}_\text{sp}$. Its $k$-th diagonal element is: 
\begin{align}
	[\mathbf{P}_\text{sp}]_{k,k}=  \exp (j 2 \pi f_\text{sp}mT_N )\exp (-j 2 \pi f_k \frac{R_\text{sp}}{c}).
\end{align}
This waveform modification with the pre-coding matrix includes one term that alters the Doppler frequency across frames, and one term that depends on the subcarrier index $k$ and distance. Due to slow-time sampling, the fake Doppler $f_\text{sp}$ in our case is captured by a single complex value unique for the $m$-th OFDM frame namely $\exp (j2\pi f_\text{sp}mT_N)$. This means that contrary to channel-induced Doppler or CFO our controlled spoofed Doppler does not differ across the samples $n$ of a single OFDM symbol (it is $e^{j2\pi f_\text{sp}n/f_s}\approx 1$) and so it does not cause inter-carrier interference (ICI). Consequently, our spoofing strategy does not incur performance loss at the nominal receiver of the OFDM signal. This result has been verified in terms of BER performance in~\cite{cnf_2023_radarconf1}.
Now the question is how to select the parameters $f_\text{sp},R_\text{sp}$ of $\mathbf{P}_\text{sp}$? First, we have to see how the receiver estimates the spectrogram.

\begin{figure*}[t]
	\centering
	\includegraphics[width=0.99\linewidth]{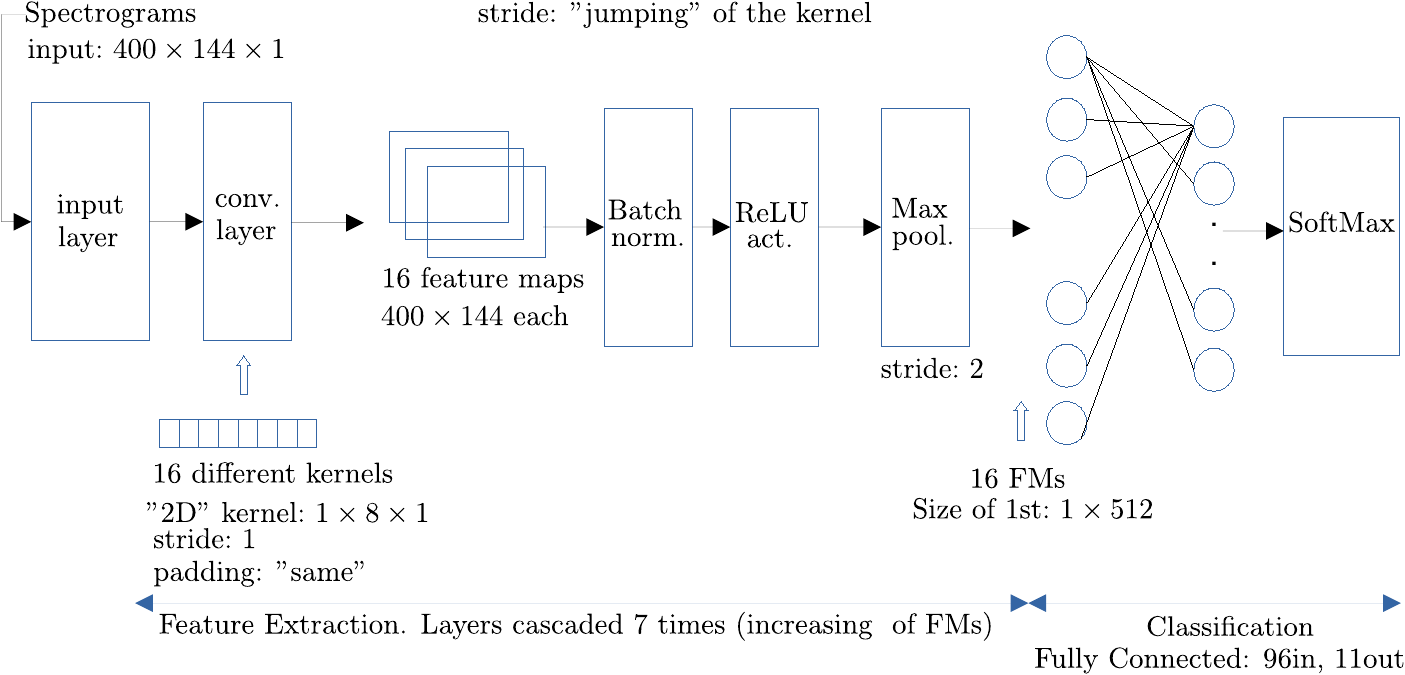}
	\caption{Input to the CNN is given as $440\times 144$ spectrograms/images. 16 feature maps are extracted over 5 convolutional layers.}
	\label{fig:cnn-for-hac}
\end{figure*}

\section{Receiver Spectrogram Calculation} 
\label{section:spectrogram-calculation}

The receiver performs a DFT on the samples $\mathbf{y}$ of all the OFDM symbols in the $m$-th frame to obtain the frequency domain (FD) signal:
\begin{align}
	\tilde{\mathbf{y}}=\mathbf{F}\mathbf{y}=\mathbf{F}\mathbf{C}_\text{cfo}\mathbf{F^{H}}(\sum_l a_l c_{\text{ch},l}\mathbf{P}_{\text{ch},l})\mathbf{P}_\text{sp}\tilde{\mathbf{x}}+\tilde{\mathbf{w}}.
\end{align}
Even though we include the matrix $\mathbf{C}_\text{cfo}$, we do not explore it, i.e., we set $\mathbf{C}_\text{cfo}$=$\mathbf{I}$. The problem of CFO in OFDM has been exhaustively investigated in other works unrelated to our problem. We now expand the last model so that it contains data from $M$ successive OFDM frames. Let $\tilde{\mathbf{X}}$ be the $N\times M$ matrix that contains these $M$ successive OFDM information blocks with QAM data $\tilde{\mathbf{x}}$. Consequently, we have $\tilde{\mathbf{Y}}$:
\begin{align}
	\tilde{\mathbf{Y}}=\mathbf{F}\mathbf{F^{H}}(\sum_l a_l c_{\text{ch},l}\mathbf{P}_{\text{ch},l})\mathbf{P}_\text{sp}\tilde{\mathbf{X}}+\tilde{\mathbf{W}}.
\end{align}
Note that $(\sum_l a_l c_{\text{ch},l}\mathbf{P}_{\text{ch},l})\mathbf{P}_\text{sp}$ is a diagonal matrix. This means that $\tilde{\mathbf{X}}$ can be removed with elementwise division as follows:
\begin{align}
	\tilde{\mathbf{Z}}=\tilde{\mathbf{Y}} \oslash \tilde{\mathbf{X}}=(\sum_l a_l c_{\text{ch},l}\mathbf{P}_{\text{ch},l})\mathbf{P}_\text{sp}+\tilde{\mathbf{W}}_f \label{eqn:Z_tilde}
\end{align}
In the above $\tilde{\mathbf{W}}_f=\tilde{\mathbf{W}} \oslash \tilde{\mathbf{X}}$. Note that the receiver uses the known preamble of the WiFi header so it knows $\tilde{\mathbf{X}}$. However, even if it does not know these symbols it can try and demodulate them (details in~\cite{cnf_2023_radarconf1}).
The $k,m$ entry of the matrix $\tilde{\mathbf{Z}}$ is:
\begin{align}
	[\tilde{\mathbf{Z}}]_{k,m}&=\sum_{l} a_l e^{-j2\pi f_k\frac{(R_l+R_\text{sp})}{c}} e^{j2\pi(f_{D_l}+f_\text{sp})mT_N}\nonumber\\
	&+W_f[k,m],0\leq k \leq N-1,0\leq m \leq M-1.	\label{eqn:signal-model-discrete-pre-process2}
\end{align}
Finally, the OFDM-based passive RADAR algorithm at the receiver performs 2D DFT on the signal in~\eqref{eqn:Z_tilde} or ~\eqref{eqn:signal-model-discrete-pre-process2} across the $k,m$ indexes with sampling periods $\Delta f$ and $T_N$ to obtain the range-Doppler response. The peak in the range-Doppler response will occur at positions $(R_l+R_{\text{sp}})/c$ and $f_{D_l}+f_{\text{sp}}$, $\forall l$. The reader is referred to~\cite{Braun14} for more details regarding this algorithm. But the essence is that with sufficient Doppler and range resolution, which for OFDM is $\Delta f$ and $1/N \Delta f$ respectively, we could make it so that the reflectors are localized in a range-Doppler response plot. As we observed in our experiments this is not possible since the range resolution of a \SI{20}{\mega\hertz} WiFi channel is very low and equal to $c/2BW\approx$ \SI{7.5}{\meter} and so we cannot distinguish anything meaningful in the range domain. Hence, for each frame $m$ we add the data in \eqref{eqn:signal-model-discrete-pre-process2} from all $N$ subcarriers. Then we calculate the spectrogram with a short-time FT (STFT) from $M$ OFDM frames.

\section{Enhancing the Attack with Randomization} With the previous analysis, we have a setup where the values for $R_\text{sp}$ and $f_\text{sp}$ are constant for the whole duration of the transmission. 
However, from the human movement model in Fig.~\ref{fig:topology-reflections} we notice that within the period that the human moves the transmitter can change multiple times the parameters of the modified signal it transmits. Consequently, we enhance our attack where every $m$-th OFDM frame changes with a different set of parameters $R_{\text{sp},m},f_{\text{sp},m}$:
\begin{align}
	[\tilde{\mathbf{Z}}]_{k,m}&=\sum_{\ell} a_l e^{-j2\pi f_k\frac{(R_l+R_{\text{sp},m})}{c}} e^{j2\pi(f_{D_l}+f_{\text{sp},m})mT_N}\nonumber\\
	&+W_f[k,m],0\leq k \leq N-1,0\leq m \leq M-1.	\label{eqn:signal-model-discrete-pre-process}
\end{align}
The spectrogram will contain frequencies at $f_{D_l}+f_{\text{sp},m}$, i.e., for all the combinations of paths and frames. In this way, we can manipulate the spectrogram more effectively.

\section{Activity Classification with Deep Learning} 
\label{section:activity-classification}

\subsection{Datasets}
To obtain our dataset we used an 802.11ac OFDM simulation system that transmits frames in the allotted bandwidth of \SI{20}{\mega\hertz}. The receiver is sampling the known preamble of the OFDM frame and produces the spectrogram as we discussed in Section~\ref{section:spectrogram-calculation}. For each scenario frames are transmitted for a duration of \SI{5}{\second}. Each human activity scenario consists of various combinations of pedestrians and bicyclists moving in a rectangular area of \SI{20}{\meter}$\times$\SI{40}{\meter}. We generate data for pedestrians that move at different speeds based on publicly available models in~\cite{Boulic90,Vishwakarma21}. These channel models generate movement of different human body parts and alter the cross-section of the human depending on the movement pattern. We also randomize in each scenario the tunable parameters for the pedestrian and cyclist objects (that are part of Matlab\textsuperscript{\tiny\textregistered}) and are presented in Tables \ref{TunPed} and \ref{TunBic}. The same number of movement scenarios were executed for each of the two signal modification strategies namely CONSTANT and RANDOM. 
The spectrograms fed into the DNN correspond to decibels but were normalized to fit in the range [0,1]. Certain signal reflections may be significantly stronger than others, so stronger signals can mask weaker ones, resulting in difficulties for analysis. To tackle this issue, logarithmic scaling (dB) is applied to enhance the features and create a fairer comparison between the received signals. In addition, amplitude normalization aids the DNN in rapid convergence.

\begin{table}[t]
	\centering
	\renewcommand{\arraystretch}{0.65}
	\caption{Tunable parameters for the pedestrian object.}
	\label{TunPed}
	\begin{tabular}{ |m{1.4cm}| >{\centering\arraybackslash}m{2.2cm} | >{\centering\arraybackslash}m{3.9cm}| }
		\hline \textbf{Parameters} & \textbf{Pedestrian} & \textbf{Acceptable values} \\
		\hline Height & $1.7 \mathrm{~m}$ & {$[1.5, 2] \mathrm{~m}$} \\
		\hline Speed & $1.3 \mathrm{~m} / \mathrm{s}$ & {$[0, 1.4 \mathrm{Height}] \mathrm{~m} / \mathrm{s}$} \\
		\hline Heading & $140^{\circ}$ & {$[-180^{\circ}, 180^{\circ}]$} \\
		\hline Location & {$[22, 4, 0] \mathrm{~m}$} & {$[[5, 45],[-10, 10], 0] \mathrm{~m}$} \\
		\hline
	\end{tabular}
\end{table}
\begin{table}[t]
	\centering
	\renewcommand{\arraystretch}{0.65}
	\caption{Tunable parameters for the bicyclist object.}
	\label{TunBic}
	\begin{tabular}{ |m{2.8cm}|>{\centering\arraybackslash}m{1.6cm} | >{\centering\arraybackslash}m{3.2cm}| }
		\hline \textbf{Parameters} & \textbf{Bicyclist} & \textbf{Acceptable values} \\
		\hline Speed & $4.5 \mathrm{~m} / \mathrm{s}$ & {$[1, 10] \mathrm{~m} / \mathrm{s}$} \\
		\hline Heading & $-30^{\circ}$ & {$[-180^{\circ}, 180^{\circ}]$} \\
		\hline Location & {$[10, -4, 0] \mathrm{~m}$} & {$[[5, 45],[-10, 10], 0] \mathrm{~m}$} \\
		\hline Gear Transmission Ratio & 4 & {$[0.5, 6]$} \\
		\hline Pedaling or Coasting & Pedaling & $50\%$ Pedaling, $50\%$ Coasting \\
		\hline
	\end{tabular}
\end{table}

\subsection{Deep Convolutional Neural Network (CNN)}
A deep CNN was chosen due to its superior performance in image classification. The architecture is illustrated in Fig.~\ref{fig:cnn-for-hac}. 
The CNN consists of five convolutional layers, followed by pooling layers, a fully connected layer, and a softmax layer. The first four convolutional layers are succeeded by a batch normalization layer, a rectified linear unit (ReLU) activation layer, and a max pooling layer. In the final convolutional layer, the max pooling layer is substituted with an average pooling layer. The output layer is a classification layer that uses softmax activation.

%
Stochastic Gradient Decent (SGD) with a mini-batch size equal to 256 was used for training. Regarding the learning rate, after several attempts to find the ideal value for the model, we concluded that it is approximately 0.02. After surpassing the 9th epoch, the learning rate is reduced by one-tenth, and so overfitting is reduced. 

\section{Performance Evaluation}
\label{section:performance-evaluation}
Using the previous datasets, and the trained classifier we evaluated its performance against our attack.
\begin{figure}[t]
	\centering
	\includegraphics[width=0.89\linewidth]{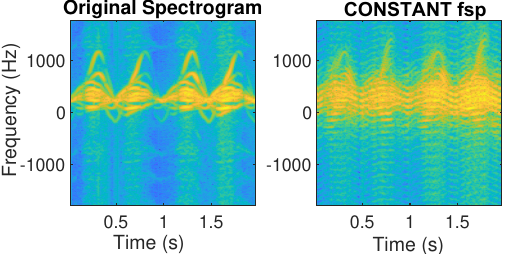}
	\includegraphics[width=0.5\linewidth]{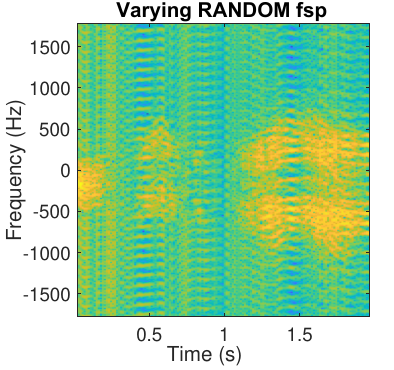}
	\caption{The spectrograms of a walking pedestrian for 3 cases: Without waveform modification (top left), waveform modified with CONSTANT range and Doppler (top right), RANDOM Doppler and range that vary across OFDM frames (bottom).}
	\label{fig:spectrograms}
\end{figure}
Regarding the selection of parameters $f_\text{sp}$ and $R_\text{sp}$ for the CONSTANT scheme we noticed that it is enough to set them to a value of \SI{100}{\hertz} and \SI{50}{\meter} respectively. For the RANDOM scheme these parameters were selected from a uniform distribution in the range \SI{50}-\SI{200}{\hertz} and \SI{10}-\SI{200}{\meter} respectively.

\textbf{Spectrograms:} 
The resulting spectrograms with the two waveform modification methods can be seen in Fig.~\ref{fig:spectrograms}. It is evident that random variation of the modified waveform across frames affects the resulting spectrograms more over time and is spread more in the frequency domain. The effect is also very satisfactory for the case of a CONSTANT frequency $f_\text{sp}$ (top right), but not as in the RANDOM where Doppler and range parameters vary across OFDM frames.

\begin{table}[b]
	\centering
	\renewcommand{\arraystretch}{0.75}
	\caption{The resulting predictions of the confusion matrix without waveform manipulation.}
	\begin{tabular}{|l|c|l|c|}
		\hline\textbf{Scenario} & Accuracy & \textbf{Scenario} & Accuracy\\
		\hline Pedestrian & $76 \%$ & Pedestr. + Bicycl. & $90 \%$ \\
		\hline Bicyclist & $98 \%$ & Pedestr. + Pedestr. & $87.5 \%$ \\ 
		\hline Bicycl. + Bicycl. & $91 \%$ & Overall & $\mathbf{88.5} \%$ \\
		\hline
	\end{tabular}
	\label{table:ResDec}
\end{table}


\begin{figure}[t]
	\centering
	\subfigure[RANDOM scheme]{\includegraphics[width=0.93\linewidth]{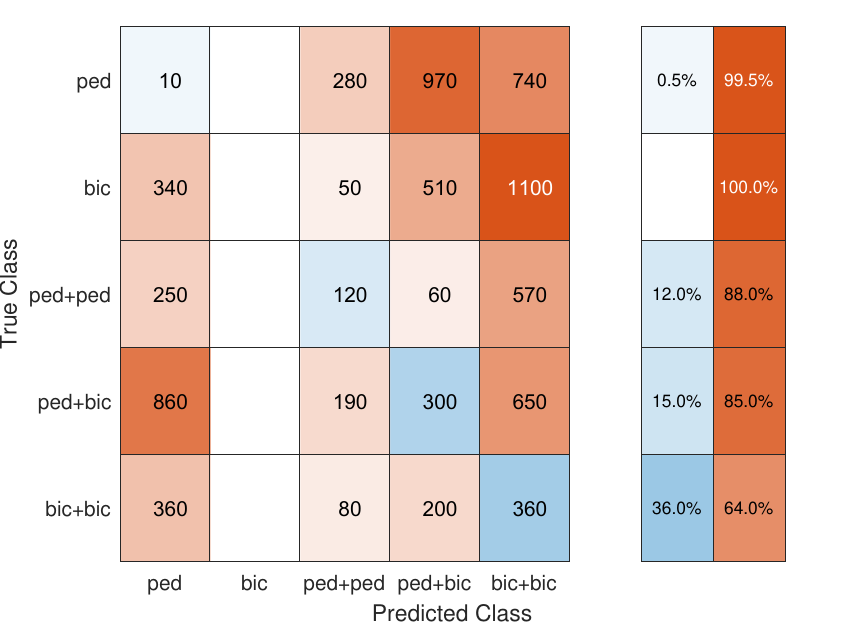}\label{fig:ConfusionMatrixULAPassive-5.11}}
	\subfigure[CONSTANT scheme.]{\includegraphics[width=0.93\linewidth]{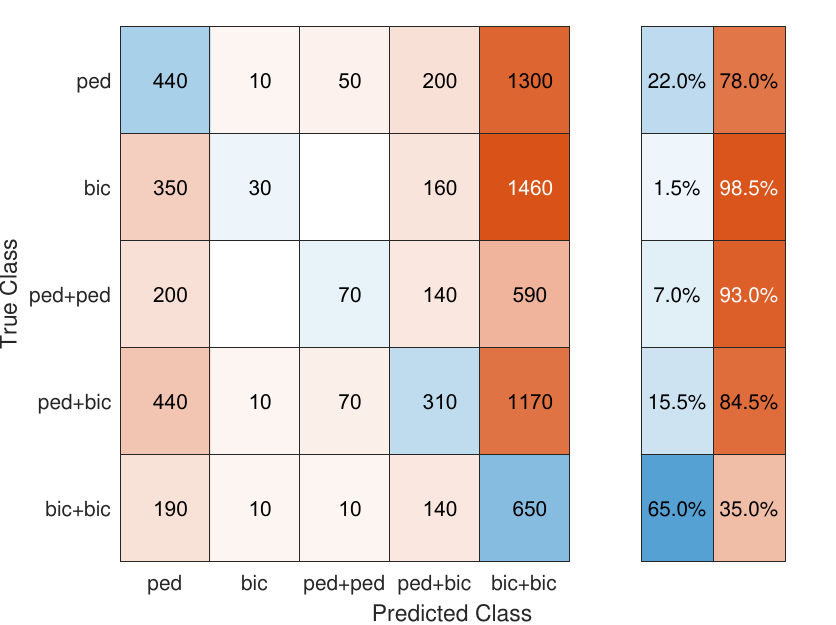}\label{fig:ConfusionMatrixULAActive-5.12}}
	\caption{Classification results with the proposed attack.}

\end{figure}

\textbf{Classification with No Attack:} 
To obtain a baseline performance we present in Table~\ref{table:ResDec} the HAC results when no attack is employed. Most prediction errors occur when the network classifies the "pedestrian" object as "pedestrian + pedestrian" or "pedestrian + bicyclist".

\textbf{Cassification with the two schemes:} For the RANDOM scheme results are shown in Fig.~\ref{fig:ConfusionMatrixULAPassive-5.11}. Each box in the confusion matrix contains the actual number of scenarios classified, while the two columns on the right contain in \% the correct and incorrect classifications. The classifier achieved a low accuracy for nearly every user activity. Overall, in all scenarios with multiple objects, the predictions of the HAC system are very inaccurate. The case where human activities can be classified with higher accuracy is the combination of two bicyclists (bic+bic). The results indicate that the DNN fails in scenarios involving a single pedestrian or a bicyclist, accomplishing a perfect attack on the classifier. It is apparent how effective this technique is in preventing HAC by concealing micro-Doppler signatures. 
%
%
The predictions of the CNN at the receiver with the CONSTANT scheme are illustrated in a more succinct form in Fig.~\ref{fig:ConfusionMatrixULAActive-5.12}, and exhibit a higher prediction accuracy over nearly all classes. HAC accuracy is slightly better indicating that this technique offers good performance but is sub-optimal compared to RANDOM. There is a minor difference in favor of this scheme when the class to be predicted is a combination of two pedestrians.



\section{Conclusions}
\label{section:conclusions}
In this paper, we presented a new attack on wireless AI-based human activity classification in systems that leverage opportunistically received wireless OFDM signals. Our method comprises the insertion of artificial frequency variations in the transmitted signal that result in the smearing of the micro-Doppler structure of the received signal when it bounces off a target. We investigated two variants for modifying the transmitted waveform at different time scales and showed their effectiveness in drastically reducing HAC accuracy. 

The paper serves a dual purpose: First, to expose the risks that HAC faces when using opportunistically received wireless signals since these can be manipulated by adversaries. Second, when the adversary now is the wireless receiver that uses an unauthorized HAC system, the proposed system can be used to protect the privacy of the targets.



\bibliographystyle{IEEEtran}
\bibliography{../../../tony-bib,../../../MyLibrary}

\end{document}